\documentclass[% 
aip,
rsi,amsmath,amssymb,
reprint,%
]{revtex4-1}

\usepackage{graphicx}% Include figure files
\usepackage{dcolumn}% Align table columns on decimal point
\usepackage{bm}% bold math
\usepackage[utf8]{inputenc}
\usepackage[T1]{fontenc}
\usepackage{mathptmx}

\usepackage{color}

\begin{document}
%\linenumbers

%\preprint{AIP/123-QED}

\title[]{Potential for improving the local realization of coordinated universal time with a convolutional neural network}
% Force line breaks with \\

\author{Takehiko Tanabe}
\email{t.tanabe@aist.go.jp}
\author{Jiaxing Ye}
\author{Tomonari Suzuyama}
\author{Takumi Kobayashi}
\author{Yu Yamaguchi}
\author{Masami Yasuda}
\affiliation{National Metrology Institute of Japan (NMIJ), 
National Institute of Advanced Industrial Science and Technology (AIST), 
1-1-1 Umeaono, Tsukuba, Ibaraki 305-8563, Japan}
\date{\today}

\begin{abstract} 
The time difference between coordinated universal time (UTC) and a hydrogen maser, 
which is a master oscillator for the local realization of UTC 
at the National Metrology Institute of Japan (NMIJ), 
has been predicted by using one of the deep learning techniques called a one-dimensional convolutional neural network (1D-CNN). 
%
% We have found that the present 1D-CNN shows better performance as a predictor 
% compared to that based on the Kalman filter. 
%
Regarding the prediction result obtained by the 1D-CNN, 
we have observed the improvement in the accuracy of prediction compared with that obtained by the Kalman filter.
Although more investigations are required to conclude that the 1D-CNN can work as a good predictor,
the present results suggest that the computational approach based on the deep learning technique 
may become a versatile method for improving the synchronous accuracy of UTC(NMIJ) relative to UTC. 
%
% A new computational approach presented in this paper may be the useful and general method 
% to improve the synchronous accuracy of UTC and its local realizations 
% for the situations in which cesium primary frequency standards, 
% ensemble of hydrogen masers and industrial cesium clocks are not available. 
%
\end{abstract}

\maketitle

% \begin{quotation}
% The ``lead paragraph'' is encapsulated with the \LaTeX\ 
% \verb+quotation+ environment and is formatted as a single paragraph before the first section heading. 
% (The \verb+quotation+ environment reverts to its usual meaning after the first sectioning command.) 
% Note that numbered references are allowed in the lead paragraph.
% %
% The lead paragraph will only be found in an article being prepared for the journal \textit{Chaos}.
% \end{quotation}

\section{Introduction}

Today, %an atomic time scale called 
coordinated universal time (UTC) % which is based on international atomic time (TAI), 
serves as the world's official time. 
Since the adoption of UTC as the world's official time about half a century ago, 
finding ways to improve reliability and long-term stability of UTC % and the accessibility of TAI and UTC 
remain the central issues in the field of time and frequency metrology \cite{Arias2011}. 
%
% TAI and 
UTC is based on the weighted average of the readings of about 500 atomic clocks 
operated at about 80 institutes around the world, 
and it is computed monthly by the Bureau International des Poids et Mesures (BIPM) \cite{Audoin2001,BIPM2017}. 
We note here that no actual clocks keep UTC, because UTC can be computed 
once all the data have been received from the international contributors.  
In other words, UTC is ``paper'' time scale calculated just once a month at 5-day intervals 
% and it is not available in real time. 
while not available in real time. 
To 
% overcome this inconvenience and 
obtain the time whenever it is needed, 
many institutes operate atomic clocks (commercially manufactured cesium atomic clocks and hydrogen masers)  
as continuously running oscillators, i.e., flywheel oscillators, 
and generate the local realization of UTC % without interruption, 
% which is commonly 
called UTC($k$), where `$k$' denotes the institute or country.
The UTC($k$) time scales have an output in real time 
and thus function as a reference in all time dissemination services that require traceability to UTC. 
%
% Because the scale interval of UTC is closely related to the definition of the second 
% in the International System of Units (SI), 
% a UTC($k$) time scale can serve as an SI-traceable reference for both time interval and frequency measurements.

\begin{figure}
\centering
\includegraphics[width=\linewidth,bb=0 0 566 314]{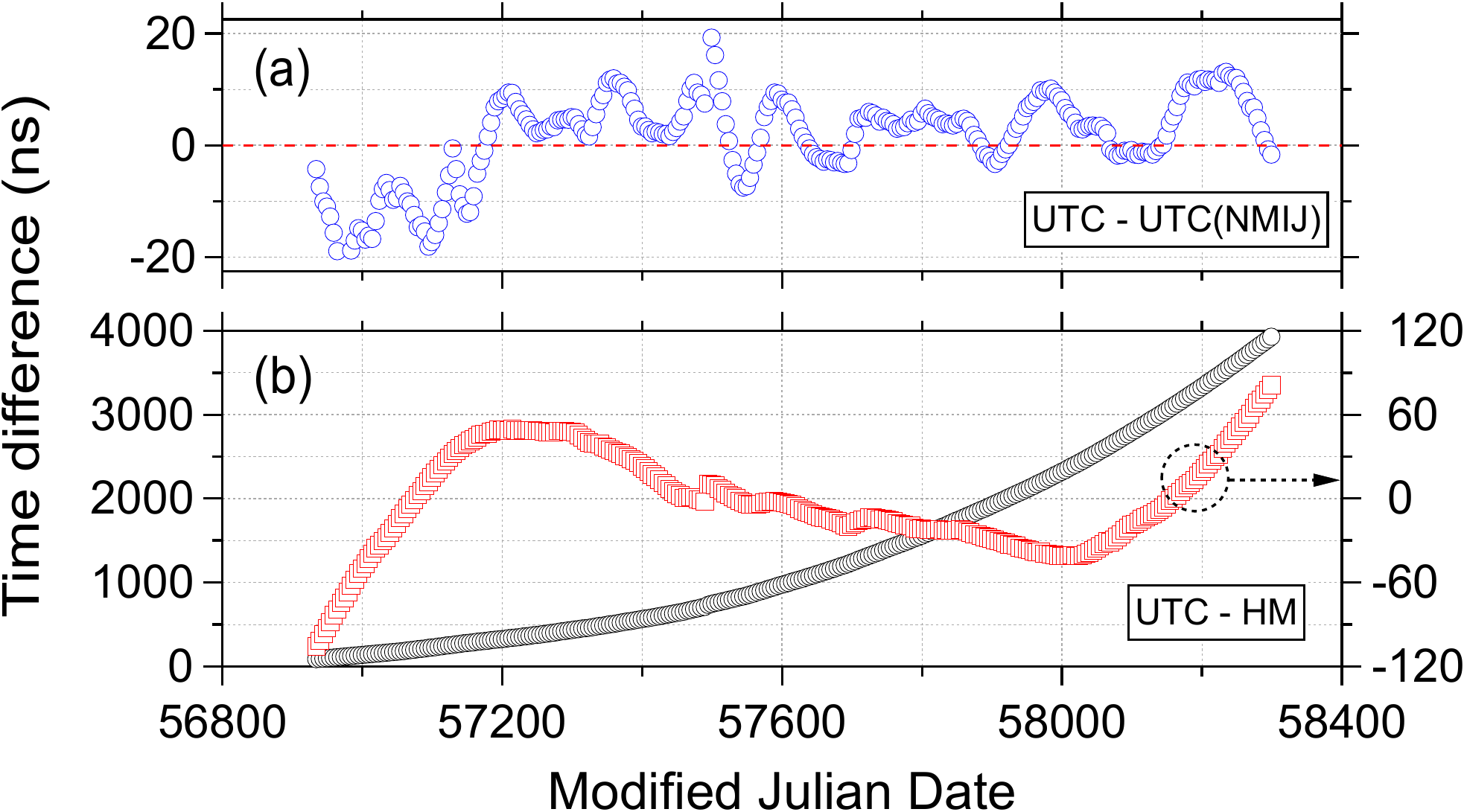}
\caption{
(a) Time difference between UTC and UTC(NMIJ) over about the last 3.5 years 
(the [UTC $-$ UTC(NMIJ)] values) reported in Circular-T. 
(b) Time difference between UTC and the master oscillator (hydrogen-maser) of UTC(NMIJ) 
(the [UTC $-$ HM] values, open circles, the left axis), and that after subtracting the quadratic component 
from the [UTC $-$ HM] values (open red squares, the right axis). 
In both figures, all vertical axes are in nanosecond units.}
\label{fig:figs1}
\end{figure}

The monthly and posterior computation of UTC yields the offsets of UTC($k$) from UTC, 
i.e., the [UTC $-$ UTC($k$)] values, 
and they are reported at 5-day intervals in the monthly document called Circular-T published by the BIPM. 
%
% According to the recommendations of the Consultative Committee for Time and Frequency, 
The offsets of UTC($k$) from UTC should be small \cite{Audoin2001}, preferably below $\pm$100 ns. 
To synchronize UTC($k$) with UTC as closely as possible, 
the frequencies of flywheel oscillators are steered by using a frequency adjuster 
based on the [UTC $-$ UTC($k$)] values. 
%
% There is no prescription for realization and keeping of UTC($k$), 
% therefore various methods to improve the synchronization accuracy of UTC and UTC($k$) are discussed \cite{Whibberley2011}. 
%
% While many local time scales are within 100 ns from UTC, 
% some of the best are within a few nanoseconds from UTC by synchronizing their flywheel oscillators on an even shorter period 
% to match the frequency of a local primary frequency standards \cite{Bauch2012}. 
%
At the National Metrology Institute of Japan (NMIJ), we maintain the UTC(NMIJ), 
which is generated using a signal from a single active hydrogen-maser (HM, Kvarz, CH1-75A) 
steered in terms of frequency by a frequency adjuster 
(the Auxiliary Output Generator (AOG), manufactured by Microsemi Corp). 
%
% The AOG generate 1 pps (pulse per second) and 5 MHz signals. 
% 
% The 5 MHz signal is frequency-doubled to 10 MHz and it is used as the frequency standard. 
% The 5 MHz signal is used as the frequency standard and UTC(NMIJ) is determined 
% by counting up the 1 pps signal from the AOG.
%
Figure \ref{fig:figs1}(a) shows the time difference between UTC and UTC(NMIJ) 
(the [UTC - UTC(NMIJ)] values), over the about last 3.5 years 
(from MJD 56934 (October 4, 2014) to MJD 58299 (June 30, 2018), MJD denotes the Modified Julian Date), 
where each point indicates the 5-day average. 
The frequency adjustment of the HM by the AOG is carried out by researchers who are well-versed in this task, 
with the consequent result that UTC(NMIJ) is within 20 ns of UTC. 
% locates at approximately $\pm$10 ns from UTC. 
% remains within about $\pm$10 ns from UTC. 

In this paper, we discuss the potential for improving the synchronous accuracy of UTC(NMIJ) 
relative to UTC by using deep learning. 
In recent years, deep learning garners plenty of research interests 
due to its superior ability in data modeling.
Deep learning is a subfield of machine learning and it is based on the artificial neural networks (ANNs).
As the name suggests, the basic building block of ANNs is a neuron, 
which works in a way analogous to the one in the human brain, 
i.e., when it receives input stimuli, outputs can be generated if the input exceeds the threshold. 
In deep learning, the ANNs are trained with the labeled data, 
e.g., images, audio and time series data. 
The ANNs consequently recognize the features of the data, 
which are sometimes complicated and cannot be identified by humans, 
and then the ANNs become able to classify or predict their future behavior. 
%
% The most direct and ideal traditionally used 
% approach to improve the synchronization accuracy of UTC and UTC($k$) is 
% calibration of them by cesium primary frequency standard. 
% through analysis of the f-f absorption or luminescence spectrum. 
%
% The fine structure in each band is ascribed to LF splitting of the ground and excited multiplets 
% and, therefore, contains direct information of the sublevel structures. 
%
Such techniques may also be applicable to the maintaining and improving UTC(NMIJ). 
If the ANNs could predict the time difference between UTC and HM precisely, 
% which is a master oscillator UTC(NMIJ), by the training using its historical time series data, 
that may help us to perform a more efficient frequency adjustment of the HM 
and accordingly provide a new and useful method for improving the synchronous accuracy of UTC(NMIJ) 
relative to UTC. 
The synchronous accuracy of UTC and UTC(NMIJ) using deep learning will be improved with the following steps, 
(i) the prediction of the time difference between UTC and HM using deep learning,
and (ii) the frequency adjustment of the HM with the AOG based on the obtained prediction results. 
As a first step, we predicted the time difference between UTC and HM using deep learning. 
%
% With a conjecture as described above, we analyzed the time difference between UTC and HM with a deep learning 
% and predicted its behavior towards the improvement of the synchronous accuracy of UTC and UTC(NMIJ). 
% using deep learning with the aim of improving the synchronous accuracy 
 
%\section*{Computational method used in this study}

\section{Computational method}

Let us explain the data treated in this study and clarify the question to be solved. 
We firstly obtained the time difference between UTC and HM (the [UTC $-$ HM] values) 
as shown in Fig.~\ref{fig:figs1}(b) (open circles, the left axis) 
with the same range as the [UTC $-$ UTC(NMIJ)] values as described above, 
by summing the [UTC(NMIJ) $-$ HM] values recorded in our group and the [UTC $-$ UTC(NMIJ)] values reported in Circular-T. 
As seen in Fig.~\ref{fig:figs1}(b), the [UTC $-$ HM] values appear to have a quadratic component as a function of MJD. 
Figure \ref{fig:figs1}(b) also shows the residual component 
after subtracting the quadratic component from the [UTC $-$ HM] values (open red squares, the right axis). 
The residual component shown in Fig.~\ref{fig:figs1}(b) changes between about -100 ns and 80 ns, 
whereas UTC(NMIJ) is within 20 ns of UTC as shown in Fig.~\ref{fig:figs1}(a). 
This fact suggests that the residual component as shown in Fig.~\ref{fig:figs1}(b) is 
corrected by the frequency adjustment of the HM, 
and needs to be predicted precisely with a view to improving the synchronous accuracy of UTC(NMIJ) relative to UTC. 
We thus predicted the residual component with deep learning. 

\begin{figure}
\centering
\includegraphics[width=\columnwidth,bb=0 0 606 328]{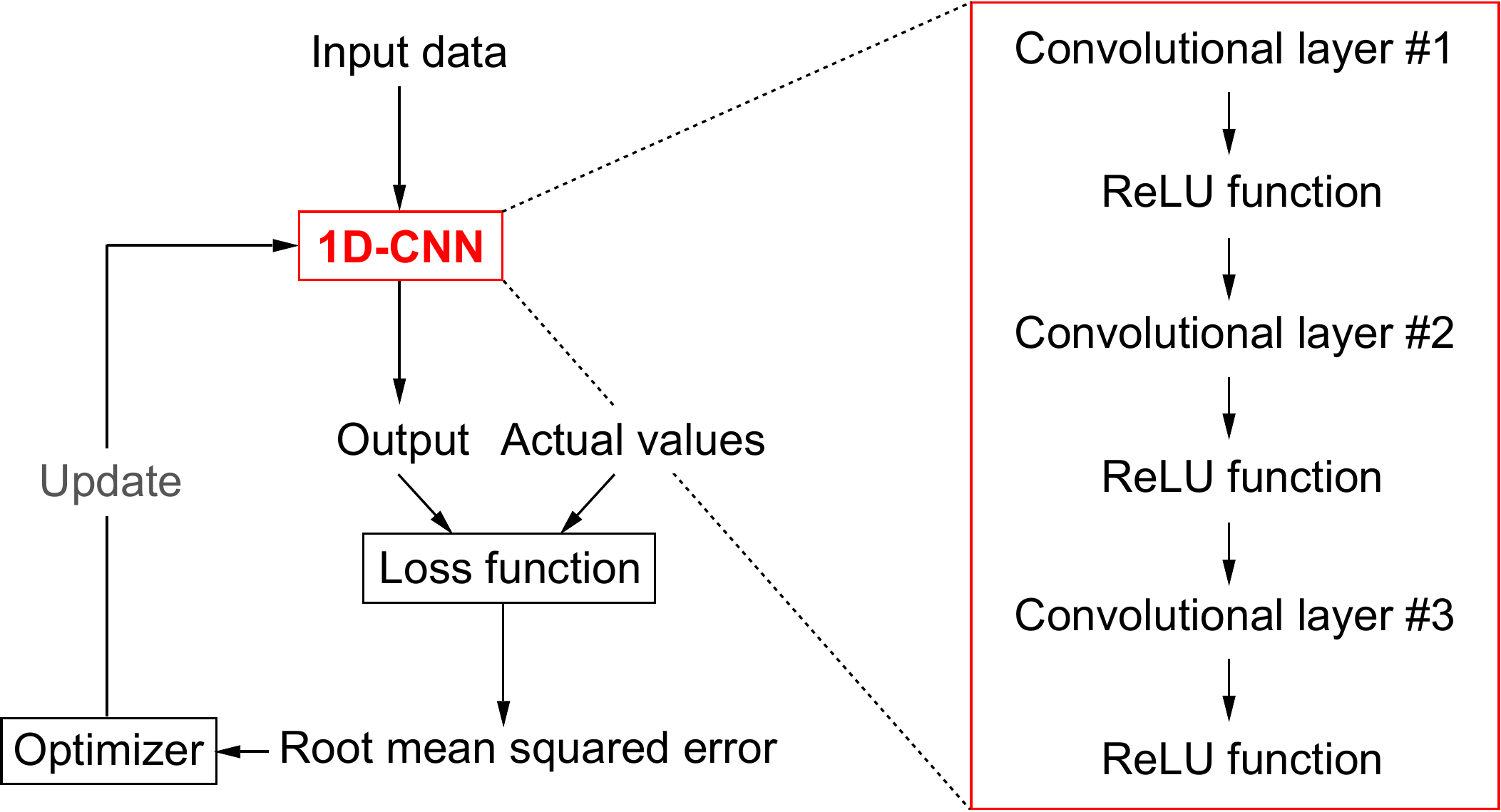}
\caption{
Training flow of the 1D-CNN and the structure of that implemented in this study. 
In the present training, the root mean squared error was calculated as the loss value 
and the Adaptive moment estimation (Adam) algorithm \cite{Goodfellow2016,Le2011} was used as the optimizer. 
All convolutional layers were followed by the Rectified Linear Unit (ReLU) function as the non-linear function. }
\label{fig:fig2}
\end{figure}

To this end, we employed a one-dimensional convolutional neural network (1D-CNN). 
CNNs are used not only for the various types of image processing \cite{Sze2017} 
but also for the analysis of the time series data as in this study \cite{Langkvist2014}. 
Although several studies on the prediction of the time difference between UTC and time scales 
using neural networks have been reported \cite{Miczulski2017,Lei2017}, 
% the ones based on the CNNs have yet to be reported. 
the effectiveness of the CNNs have not been explored. 
Figure \ref{fig:fig2} shows the training flow of the 1D-CNN 
and the structure of that implemented in this study. 
The CNNs are essentially a stack of the convolutional layers, which are composed of plural neurons,  
and the part of neurons is coupled to the ones in another convolutional layer with certain weights. 
In the convolutional layers, the features of the data are recognized 
% where the data are treated as the matrix, 
% by calculating the element-wise multiplication and addition of the data and the filters. 
by using the filters. 
%
% where the all pixels of the image are digitized and treated as the matrix, 
% by calculating the element-wise multiplication and addition of the image and the filters (matrices) 
% with sliding them over the image. 
%
% It is a mathematical operation that takes two inputs such as image matrix and a filter or kernal
%
% Subsequent to the convolutional layer, it is usual to put the non-linear layer.
% The purpose of the non-linear layers is to introduce the non-linearity into the network. 
%
The prototype convolutional layers are linear systems, because their outputs constitute the 
multiplication and addition of the input data and the filters. 
To enhance the expressiveness of the CNNs the non-linear 
% layers 
activation functions are introduced following the convolutional layers. 
The goal of deep learning is to generalize the ANNs, 
that is, to realize ANNs with good performance even for data it has never seen before. 
% 
% It is therefore important to measure the degree of generalization of the ANNs. 
%
% In deep learning, the ANNs is able to learn and automatically distill information 
% from the input data and such ability is gained through the training process of the ANNs. 
%
For this purpose, the loss values, 
i.e., the differences between the outputs of the ANN and their corresponding actual values, 
are calculated by using the loss function in the training process. %more detail, 
%
% training means finding a set of parameters of the networks that minimize the loss 
% for a given set of training data and their corresponding target values. 
%
The ANNs including CNNs are basically a composition of multiple functions with many weights. 
The optimizer updates the weights so as to minimize the loss values.

The total number of the discrete time difference data between UTC and HM 
shown in Fig.~\ref{fig:figs1}(b) was 274 and we denote this data as $d_{i}\ (i=1, 2, ..., 274)$. 
%
% be the discrete time difference data between UTC and HM 
% shown in Fig.~\ref{fig:figs1}(b) and the total number of the data was 274. 
%
We normalized the data by the maximum of the absolute value 
such that $x_{i} = d_{i} / |d_{\text{max}}|$. 
The initial 51\% of the normalized data, noted by ${\bf x}^{\text{trn}} = [x_{1},\ x_{2},\ ...\ ,x_{141}]$, 
were fed into the 1D-CNN for the training. 
The features of the input data were recognized in the three convolutional layers
by using three filters of which size were 1 $\times$ 4, 1 $\times$ 3 and 1 $\times$ 3. 
All convolutional layers were followed by the Rectified Linear Unit (ReLU) functions \cite{Goodfellow2016}, 
which returns zero if the input values are zero or less, 
but it returns that values for positive input values. 
The ReLU function induces non-linearity and sparsity for better training \cite{Glorot2011}. 
The root mean squared error $E_{\text{RMS}}^{\text{trn}}$ was calculated as the loss value 
in the present training, which is defined as 
\begin{equation}
E_{\text{RMS}}^{\text{trn}} = \sqrt{\frac{1}{n_{\text{trn}}}\ {\displaystyle \sum_{i=1}^{n_{\text{trn}}} \left( x_{i}^{\text{pred}} - x_{i} \right)^{2}}},\ % i \in [1,\ n_{\text{trn}}],
\end{equation}
where $n_{\text{trn}}$, $x_{i}^{\text{pred}}$ are the number of the training data, 
$i$-th prediction by the 1D-CNN in the training, 
such that $x_{i}^{\text{pred}} = f({\bf x}^{\text{trn}}, {\bf W})$, 
and {\bf W} is the weights of the 1D-CNN to be optimized in the training. 
From the comparison between some optimizers, 
the Adaptive moment estimation (Adam) algorithm \cite{Goodfellow2016} was selected as the optimizer.
The Adam algorithm is widely used in the CNNs due to its high-efficiency and low computational cost \cite{Le2011}. 
%
% The Adam algorithm is one of the stochastic gradient descent for the optimization for the functions. 
%
% The Adam algorithm is an advanced method which renders multiple favorable properties, 
% including high computational efficiency, well suited for problems with massive data or parameters, 
% and low memory requirement 
%
% In the training process, it is necessary to measure how far the output of the 1D-CNN are 
% from their corresponding actual values, and this is the job of the loss function. 
%
% and only the optimal settings are shown in the manuscript.
%
The computer program was written with MATLAB programming language \cite{MATLAB2018}.
%
% TensorFlow is a software library for machine learning, 
% and Keras is a wrapper library that uses TensorFlow as its back-end engine. 
%
% Keras is an open source python library that enables you to easily build Neural Networks. 
% The library is capable of running on top of TensorFlow. 
% Both TensorFlow and Keras are software libraries for machine learning and 
% Keras is a library for deep learning that can run on top of TensorFlow.

% The learning rate is set to $10^{-4}$, and the other parameters are set to the default values 
% recommended by the authors of the Adam algorithm. 
% We also utilize two dropout layers in the middle of the neural network architecture 
% to regularize the training to avoid overfitting the data. 

% Although we tried training several models, and specifically investigated the effect of image padding
% (including extra space around the images), the accuracy changed little because our network has only one
% convolutional layer and the detailed features around the image boundary are not so important for the four basal
% shapes. 

% The pooling layer, which removes the effect of trivial movement of objects in images and increases
% the robustness of the network to displacements of objects, has a stride equal to the length of the filter and selects
% the maximum value in each area ($2 \times 2$ block). 
% 
% After the pooling, the signals are sent to an affine layer, and then
% the probabilities for each of the four classes are calculated. 

\section{Results and discussions}

\subsection{Results of training}

\begin{figure}
\centering
\includegraphics[scale=0.4,bb=0 0 477 388]{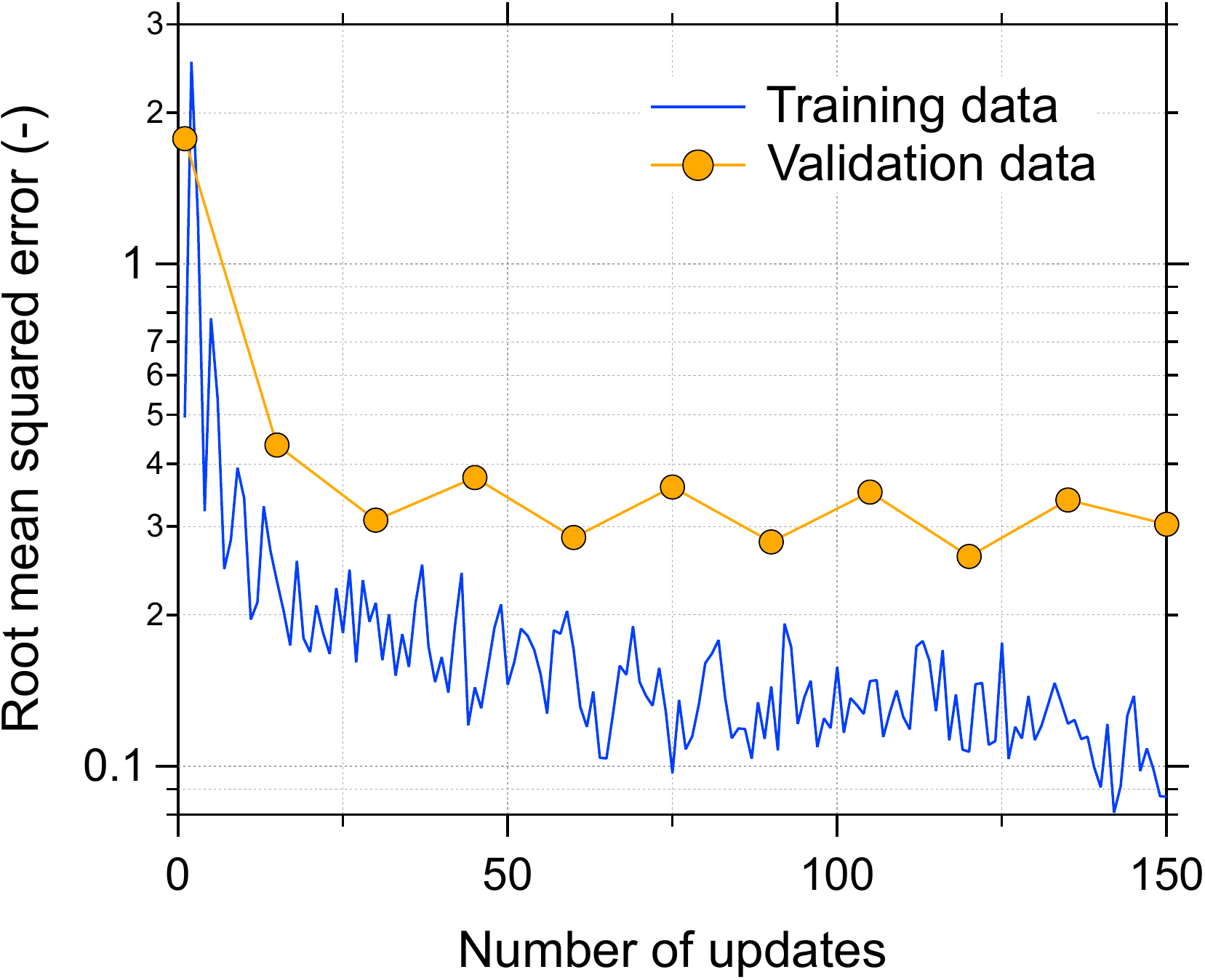}
\caption{
The results of the training, 
i.e., the root mean squared error of the training data and that of the validation data 
as a function of the number of weights updates with a logarithmic scale of the vertical axis.}
\label{fig:fig3}
\end{figure}

Figure \ref{fig:fig3} shows the results of the training of the 1D-CNN, 
i.e., $E_{\text{RMS}}^{\text{trn}}$ of the training data 
and that of the validation data as a function of the number of weights updates 
with a logarithmic scale of the vertical axis. 
$E_{\text{RMS}}^{\text{trn}}$ of the training data gradually decreased 
and changed slightly after about the 50th update. 
This indicates that the weights of the 1D-CNN were optimized to the training data 
after about the 50th update. 
In the present study, the 12\% of normalized data (33) were used as the validation data, 
that were not used in the training. 
The training of the 1D-CNN involves some parameters, 
e.g., the number of the convolutional layers and filters, the types of non-linear functions and optimizers. 
All of these parameters were determined so as to converge 
$E_{\text{RMS}}^{\text{trn}}$ of the validation data. 
As for $E_{\text{RMS}}^{\text{trn}}$ of the validation data shown in Fig.~\ref{fig:fig3}, 
similar behavior to that of the training data was observed 
after about the 25th update. 
One point to be noted as regards the training is the overfitting \cite{Goodfellow2016}, 
which happens when the ANNs are tuned only for the training data and do not generalize to unseen data. 
Once overfitting occurs, $E_{\text{RMS}}^{\text{trn}}$ of the validation data 
is expected to exhibit monotonically increasing trend starting from a certain update. 
Although there are some method to avoid the overfitting, 
we employed the L2 regularization and the early stopping method \cite{Goodfellow2016}. 
% % 
% By introducing the early stopping, the training of the 1D-CNN was stopped 
% when the $E_{\text{RMS}}^{\text{trn}}$ of the validation data begins to exhibit the monotonically increasing trend. 
% }
$E_{\text{RMS}}^{\text{trn}}$ of the validation data shown in Fig.~\ref{fig:fig3} indicates that 
overfitting has not occurred in the present 1D-CNN. 
From the above, it is reasonable to consider that the training of the 1D-CNN had been conducted properly. 
%
% That's what you would expect when running gradient descent optimization 
% the quantity you're trying to minimize should be less with every iteration.
%
% As the epoch number increases, the loss functions using mini-batch decrease and the training and test accuracies
% increase. 
% From the 80th epoch, the loss function becomes stable, so we decided to run 500 epochs. 
% During training, the training loss is higher than the testing loss. 
% This is because the regularization mechanism of the dropout layer is turned off at testing time. 
% We confirmed that the training loss and the testing loss became the same if the two dropout layers are removed. 
% 
% The accuracy of the CNN after 100 epochs was evaluated from test images to be approximately 92\%. 
%
% Because initial values of layers and the mini-batch were determined randomly, 
% we trained the CNN several times, and confirmed that the corresponding changes in accuracy were negligible.

\begin{figure}
\centering
\includegraphics[width=\linewidth,bb=0 0 749 544]{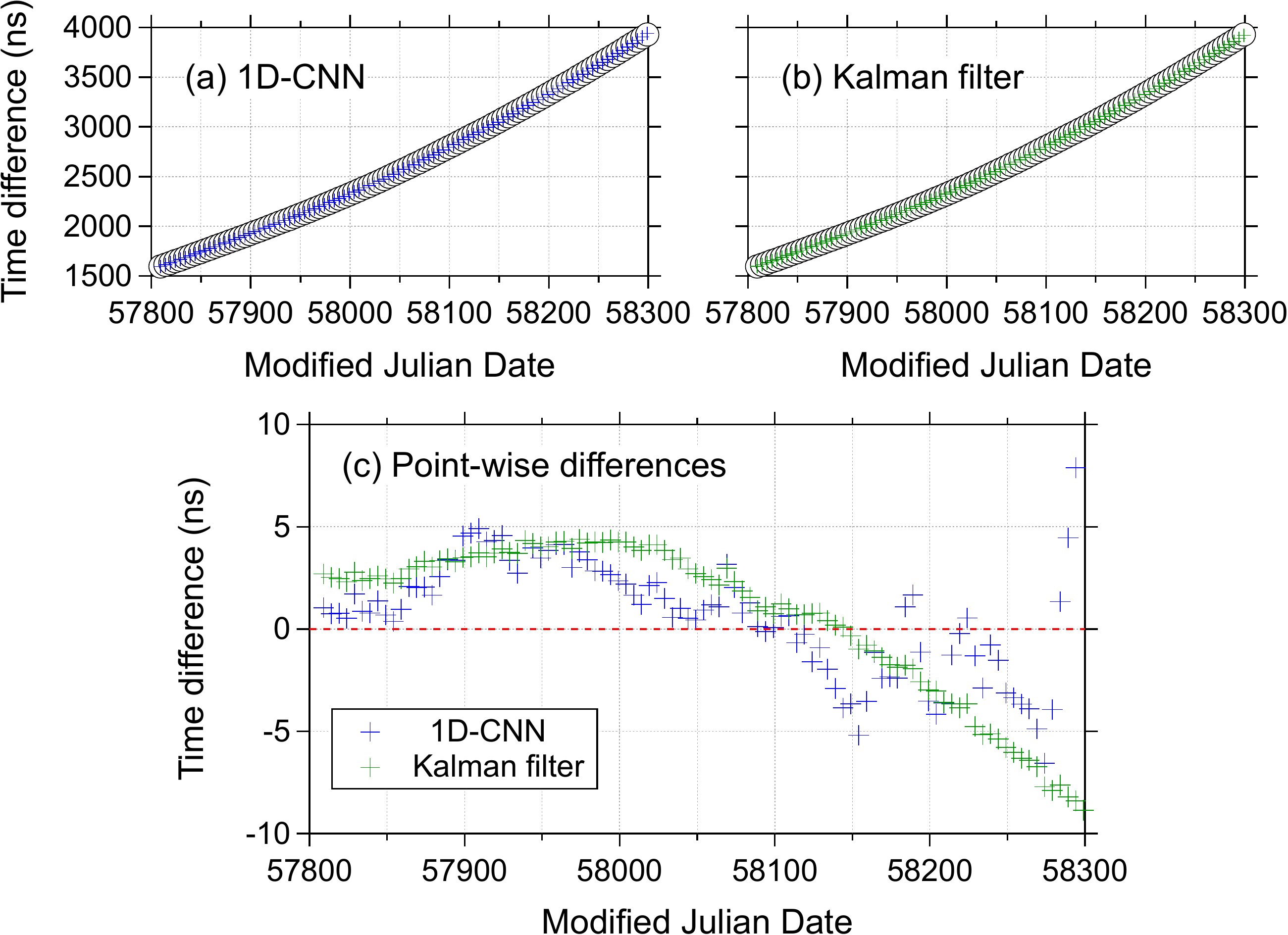}
\caption{
Typical prediction result obtained by the 1D-CNN (a) and the Kalman filter (b). 
In both Figs. (a) and (b), the test data are shown as the open circles. 
The point-wise differences between the test data and the prediction results obtained by two methods are shown in (c).
}
\label{fig:figs4}
\end{figure}

\subsection{Prediction results and discussions}

The typical prediction result obtained by the trained 1D-CNN is shown in Fig.~\ref{fig:figs4}(a) (blue crosses). 
In Fig.~\ref{fig:figs4}(a), the actual data (open circles), 
which were used as neither the training nor the validation data, 
are also shown and we hereinafter refer to this data as the test data. % $x_{i}^{\text{test}}$. 
Figure~\ref{fig:figs4}(b) shows the prediction by the Kalman filter (green crosses) as mentioned later 
and Fig.~\ref{fig:figs4}(c) shows the point-wise differences 
between the test data and the predictions obtained by two methods. 
The predictions by the 1D-CNN and the Kalman filter were performed 
by repeating the short-term prediction over the whole data as follows; 
5 points of test data (open red circles in Fig.~{\ref{fig:figs1}}(b)) 
were fed into the 1D-CNN and the Kalman filter and they predicted one point ahead. 
Both prediction results shown in Figs.~\ref{fig:figs4}(a) and (b) were eventually obtained 
by summing the predicted residual components and the subtracted quadratic component. 
The procedure for the prediction as described above means that 
the results of the previous predictions were not taken into account in each prediction cycle. 
% It means that the results of the previous predictions were not taken into account in each prediction cycle. 
%
In this sense, we have to admit that the performance of the present 1D-CNN as a predictor 
is at an early stage. 
%
% \sout{
% Although we have changed the parameters of the present 1D-CNN, 
% e.g., the numbers of the layers and filters, % and updates of their weights, 
% and the types of non-linear functions and optimizers, we observed no obvious changes in the prediction result. 
% }
%
% \sout{
% This may be linked to the fact that there was insufficient training data (160) 
% compared with the number of the parameters to be optimized (4417) in the 1D-CNN, 
% and it is difficult to resolve this immediately.  
% }
%
% It implies that the neural network-based predictor is working well 
% and that the behavior of maser are predictable in the long term. 
%
%
% In addition, Besides the above, 
% 
As can be seen from Fig.~\ref{fig:figs4}(c), 
the point-wise differences between the test data and the predictions 
by two methods tended to increase in the latter half of the predictions. 
This may be related to the environmental variations of the room where the HM is located. 
In fact, the room temperature of the range from MJD 58200 to 58300 significantly fluctuated 
compared to the other range. 
This may resulted in the discrepancy between the test data and the prediction results. 
% shown in the range between MJD 58200 to 58300. 
%
In other words, the prediction may be improved 
by considering the environmental variations of the room where the HM is located.

As described above, we also performed the prediction by using the Kalman filter 
% evaluated the prediction result obtained by the 1D-CNN by comparing that obtained 
% by the Kalman filter 
on the same training and test data.
The Kalman filter is a linear iterative method for modeling the continuous observables 
affected by the random noise \cite{Harvey1990}. 
%
% The Kalman filter is a linear iterative approach to model sequential observations over time 
% containing noises by using a joint probability distribution \cite{Harvey1990}. 
% based on which it produces estimates that tend to be more accurate than those 
% based on a single measurement alone . 
%
% The prediction by the Kalman filter was performed in the same manner as that of the 1D-CNN. 
%
There exist some studies on the prediction of the time difference between UTC and the atomic clocks 
with the Kalman filter \cite{Davis2011}. 
Let us calculate the root mean squared error of the results of the predictions $E_{\text{RMS}}^{\text{pred}}$. 
$E_{\text{RMS}}^{\text{pred}}$ is defined as  
\begin{equation}
E_{\text{RMS}}^{\text{pred}} = \sqrt{\frac{1}{n_{\text{pred}}}\ 
{\displaystyle \sum_{j=1}^{n_{\text{pred}}} \left( x_{j}^{\text{pred}} - x_{j}^{\text{test}} \right)^{2}}},\ % j \in [1,\ n_{\text{pred}}], 
\end{equation}
where $n_{\text{pred}}$, $x_{j}^{\text{pred}}$ and $x_{j}^{\text{test}}$ are the number of the predicted data, 
the results of predictions and the test data, respectively. 
%
% In this evaluation, the $E_{\text{RMS}}^{\text{pred}}$ of the prediction 
% obtained by the Kalman filter is anticipated to be higher than that of the 1D-CNN. 
%
$E_{\text{RMS}}^{\text{pred}}$ of the predictions obtained by two methods shown in Figs.~\ref{fig:figs4}
were calculated to be, 
\begin{equation*}
\text{1D-CNN : } \approx 3.0\ \text{ns},\ \text{Kalman filter : } \approx 3.8\ \text{ns}.
\end{equation*}
%
% We observed improvement in the $E_{\text{RMS}}^{p}$ of the prediction obtained by the 1D-CNN 
% by a factor of about 1.3 compared with that obtained by the Kalman filter. 
%
In the results of repeating the predictions, 
we observed improvement in $E_{\text{RMS}}^{\text{pred}}$ of the prediction 
obtained by the 1D-CNN compared with that obtained by the Kalman filter. 
%
% \textcolor{blue}
% {
% The achieved improvement by the 1D-CNN is considered to be obtained for the following reason; 
% while the Kalman filter is grounded on the Gaussian model assumption, 
% which can be constrained so as to describe the actual data, the 1D-CNN exploit larger hypothesis space 
% than that of the Kalman filter. 
% }
%
%
The 1D-CNN has been designated to exploit complex dependencies 
among input variables via many layers of non-linear operators for the prediction. 
On the other hand, the Kalman filter recursively performs a conditional probability estimation, 
and this method is commonly anticipated to be optimal under the Gaussian model assumption. 
%
% if the errors are Gaussian distribution. 
%
In other words, the 1D-CNN exploited larger hypothesis space than that of the Kalman filter for the prediction. 
The improvement in $E_{\text{RMS}}^{\text{pred}}$ of the prediction by the 1D-CNN as shown above 
% is considered to be obtained by 
can be attributed to the higher expressiveness of the 1D-CNN than that of the Kalman filter, 
i.e., high non-linearity and superior ability to exploit complex relationships between input variables. 
Furthermore, the difference in the mathematical assumptions between two methods as described above 
also % appeared to be 
reflected in the error trend in Fig.~\ref{fig:figs4}(c); 
%
% We can find more detailed differences in the error trends between the two prediction results; 
while the prediction result obtained by the 1D-CNN (blue crosses in Fig.~\ref{fig:figs4}(c)) 
was more wiggling than that obtained by the Kalman filter, 
the prediction result obtained by the Kalman filter (green crosses in Fig.~\ref{fig:figs4}(c)) 
exhibits a mostly continuous and monotonic curve, 
which is consistent with recursive model updating algorithm of the Kalman filter. 
%
% that obtained by the Kalman filter (green crosses in Fig.~\ref{fig:figs4}(c)) was mostly continuous and monotonic. 
%
% This difference is also considered to be the by two methods;
%
% The wiggling in the prediction result obtained by the 1D-CNN 
% can also be attributed to the high non-linearity of the 1D-CNN. 
%
% This fact is consistent with the difference in the mathematical assumptions 
% between two methods as described above.
%
%
Although more investigations are required to conclude that the 1D-CNN can work as a good predictor, 
% to establish a fine-tuned 1D-CNN that can work as the good predictor, 
%
the present results suggest that our new computational approach may accordingly provide a useful method 
for improving the synchronous accuracy of UTC(NMIJ) relative to UTC.
We are now working on the detail investigations towards establishing the reliable method 
for predicting the [UTC $-$ HM] values % by using various types of ANNs including the 1D-CNN, 
and the results will be comprehensively discussed in our forthcoming paper including the prediction 
by the Kalman filter.

\section{Conclusion}

We have predicted the time difference between UTC and HM, 
which is used as a master oscillator for UTC(NMIJ) by using a 1D-CNN, 
and observed the improvement in the accuracy of prediction compared with that obtained by the Kalman filter. 
The prediction may be improved by considering the environmental variations of the room where the HM is located. 
In addition, the present study focused on not the efficiency of the prediction 
but the accuracy of the prediction, 
therefore the present 1D-CNN has not been optimized regarding the speed of the computation. 
We will address the points as above in the near future towards establishing the deep learning based method 
for improving the synchronous accuracy of UTC(NMIJ) relative to UTC. 
It should be emphasized that the method discussed in this paper will also be available 
not only for improving the synchronous accuracy of UTC(NMIJ) relative to UTC but also for other UTC($k$) time scales. 
Such versatility and application potential attract much interests. 

\nocite{*}
%\bibliographystyle{junsrt}
%\bibliography{References_byTanabe}

%

\end{document}